\documentclass[a4paper,10.5pt]{article}
\usepackage{amssymb}
\usepackage{amsmath}

\usepackage{latexsym}
\usepackage{array}
\usepackage{bm}
\usepackage{theorem}
\usepackage{graphicx}
\usepackage{exscale}

\theoremstyle{break}
\newtheorem{Theorem}{Theorem}[section]
\newtheorem{Proposition}[Theorem]{Proposition}
\newtheorem{Lemma}[Theorem]{Lemma}
\newtheorem{Corollary}[Theorem]{Corollary}

\def\Proof{\hfil\break{\bf Proof}\;\;\;\;}

\def\hbreak{\vspace*{2mm}\hfill\break\noindent}

\def\p{\boldsymbol{p}}
\def\q{\boldsymbol{q}}

\def\b0{\boldsymbol{0}}

\def\Z{\mathbb{Z}}

\def\qed{\hfill\hbox{\rule[-2pt]{3pt}{6pt}}}

\def\det{\mbox{det}\,}

\begin{document}

\title
{A two-dimensional lattice equation as an extension of the Heideman-Hogan recurrence} 
\author
{Ryo Kamiya$^1$, Masataka Kanki$^2$, Takafumi Mase$^1$, Tetsuji Tokihiro$^1$\\
\small $^1$ Graduate School of Mathematical Sciences,\\
\small the University of Tokyo, 3-8-1 Komaba, Tokyo 153-8914, Japan\\
\small $^2$ Faculty of Engineering Science,\\
\small Kansai University, 3-3-35 Yamate, Osaka 564-8680, Japan}

\date{}

\maketitle

\begin{abstract}
We consider a two dimensional extension of the so-called linearizable mappings.
In particular, we start from the Heideman-Hogan recurrence, which is known as one of the linearizable Somos-like recurrences, and introduce one of its two dimensional extensions. The two dimensional lattice equation we present is linearizable in both directions, and has the Laurent and the coprimeness properties. Moreover, its reduction produces a generalized family of the Heideman-Hogan recurrence. Higher order examples of two dimensional linearizable lattice equations related to the Dana-Scott recurrence are also discussed.
\end{abstract}

\section{Introduction}
For rational second order mappings, there are two important criteria of integrability: i.e., the singularity confinement\cite{SC} and zero algebraic entropy\cite{HV}. The former has been successfully utilized to obtain a series of discrete Painlev\'{e} equations\cite{RGH}
 and the latter has provided a keen and handy tool for distinguishing integrable mappings from non-integrable ones\cite{BV}.
Although typical second order `integrable' mappings, such as the discrete Painlev\'{e} equations and the QRT mappings, satisfy both criteria,
there are exceptions that meet only one of them.
The celebrated Hietarinta-Viallet equation is one of the mappings which have singularity confinement property but have positive algebraic entropy\cite{HV}.
Recently, Mada and three of the authors have proposed a criterion, called the `co-primeness property', which is regarded as an algebraic reinterpretation of the singularity confinement property\cite{KMMT2}.
We have constructed a two dimensional discrete lattice equation which has co-primeness property. The equation, through a reduction (i.e., a projection of the domain of definition onto a line), generates a family of nonlinear mappings with positive algebraic entropy, including the Hietarinta-Viallet equation\cite{KMT}.
Two and three-dimensional lattice equations with similar properties have also been constructed by the deformation of the discrete Toda lattice equation\cite{KKMT}.
Another class of exceptional second order mappings are so called the linearizable mappings\cite{RGSM}, which do not have the singularity confinement property but have zero algebraic entropy.
A natural question is whether we have higher dimensional lattice equations which can be regarded as the extension of the linearizable mappings.    

An example of the typical linearizable mappings is the following recurrence\cite{DS}
\begin{equation}\label{DS_rec}
a_{n+1}a_{n-3}=a_na_{n-2}+a_{n-1},
\end{equation} 
which is one of the recurrences discovered by Dana Scott to have integrality when all the initial values are chosen to be units.
Equation \eqref{DS_rec} 
has been discovered to be written down as a mutation of a quiver in the terminologies of the cluster algebras\cite{FZ}. Also \eqref{DS_rec} and its generalization have been investigated in detail with respect to Poisson structures\cite{Fordy-Hone}.
The following two dimensional system is known as the equation for the two-frieze pattern \cite{Propp,M-GOT}
\begin{equation}\label{two_frieze}
v_{i,j}=v_{i+1/2,j+1/2}v_{i-1/2,j-1/2}-v_{i+1/2,j-1/2}v_{i-1/2,j+1/2} \quad ((i,j) \in \Z^2 \mbox{or} (\Z+1/2)^2),
\end{equation} 
which can be interpreted as a two-dimensional extension of \eqref{DS_rec}.
Another interesting linearizable mapping we mainly study in this paper is the Heideman-Hogan recurrence\cite{Heideman-Hogan}:
\begin{equation}\label{HH_eq}
a_{n+2k+1}a_n=a_{n+2k}a_{n+1}+a_{n+k}+a_{n+k+1},
\end{equation}
where $k$ is a positive integer.
The recurrence \eqref{HH_eq} is one of Somos-like recurrences, since, if we assign $a_l=1 (l=0,1,\cdots, 2k)$, then $a_n\in\mathbb{Z}$ for all $n\ge 2k+1$.
It is known that the nonlinear recurrence \eqref{HH_eq} satisfies the linear equation
\begin{equation}
a_{n+6k}-K(a_{n+4k}-a_{n+2k})-a_n=0, \label{LinHHgen}
\end{equation}
where $K$ is a constant determined by the initial data $\{a_0,a_1,\cdots, a_{2k}\}$.
If all the initial values are equal to 1, then \eqref{LinHHgen} has the following simple form
\[
a_{n+6k}-(2k^2+8k+4)(a_{n+4k}-a_{n+2k})-a_n=0.
\]
The Laurent property\cite{FZ2} and explicit form of $K$ are discussed in \cite{Hone-Ward}.

In this article, we introduce a linearizable two dimensional lattice equation
\begin{equation}\label{2DHH}
x_{n+2,t+1}x_{n,t}=x_{n,t+1}x_{n+2,t}+x_{n+1,t+1}+x_{n+1,t}\quad ((n,t)\in \Z^2),
\end{equation}
the reduction of which gives the Heideman-Hogan recurrence \eqref{HH_eq} and its generalization.
In fact, when we impose $x_{n,t}=x_{n+1,t-k}$ for all the iterates of the equation \eqref{2DHH} and define $a_j=x_{n,t}$ where $j=kn+t$, then the new variable $a_j$ satisfies the original Heideman-Hogan recurrence \eqref{HH_eq}.
In the next section, we present our main theorem stating that Eq. \eqref{2DHH} is linearizable, and give its proof.
In section 3, we consider the reduction of \eqref{2DHH}. We also study the Laurent property and the degree growth of the equation \eqref{2DHH}.
We prove that the iterate $x_{n,t}$ of \eqref{2DHH} is an irreducible Laurent polynomial of the initial variables, and that every pair of two iterates are coprime as Laurent polynomials, which are indications of the integrability of the equation.
We also prove that the degree of $x_{n,t}$ grows as a polynomial of the first order.
Some extensions of \eqref{two_frieze} using matrix identities are discussed in section 4.
Section 5 is devoted to concluding remarks.
%
%
%
\section{Linearizable lattice equation related to Heideman-Hogan recurrence}
Let us consider the two-dimensional lattice equation:
\[
x_{n+2,t+1}x_{n,t}=x_{n,t+1}x_{n+2,t}+x_{n+1,t+1}+x_{n+1,t}\quad ((n,t)\in \Z^2).
\]
Here we interpret that \eqref{2DHH} determines $x_{n+2,t+1}$ from the other 5 terms.
We show that \eqref{2DHH} is a linearizable lattice equation.
The following theorem is our main result in this article.
\begin{Theorem}\label{Theorem_main}
A solution $x_{n,t}$ of \eqref{2DHH} satisfies the following linear equations: 
\begin{subequations}
\begin{align}
&x_{n+6,t}+\alpha(n)x_{n+4,t}+\beta(n)x_{n+2,t}-x_{n,t}=0,  \label{Lin1}\\
&x_{2m,t+3}+\gamma(t)x_{2m,t+2}+\delta(t)x_{2m,t+1}+\epsilon(t)x_{2m,t}=0,  \label{Lin2}\\
&x_{2m+1,t+3}+\gamma'(t)x_{2m+1,t+2}+\delta'(t)x_{2m+1,t+1}+\epsilon'(t)x_{2m+1,t}=0,  \label{Lin3}
\end{align}
\end{subequations}
where $\alpha(n), \beta(n)$ are functions of $n$, which are independent of $t$. Also $\gamma(t), \ldots, \epsilon'(t)$ are functions of $t$, which are independent of $n$.
\end{Theorem}

In order to prove Theorem \ref{Theorem_main}, we need several lemmas and propositions.
The following lemma (Dodgson condensation \cite{Dodgson}, Lewis Carrol identity, or Desnanot-Jacobi identity) is the key lemma highly used to investigate linearizable mappings\cite{Fordy-Hone, Hone-Ward}.
\begin{Lemma}\label{Dodgson}
Let $A$ be an $n \times n$ matrix, $|A_{ik}|$ be its $(ik)$ minor
(determinant of $A_{ik}$, which is obtained by deleting the $i$th row and the $k$th column from $A$), and $|A_{ij;kl}|$ be its $(n-2)\times(n-2)$ minor which is obtained by deleting the $i$th and the $j$th rows and the $k$th and the $l$th columns. Then we have
\begin{equation}\label{Dodgson_id}
\left| A_{1n;1n}\right| \left| A\right|=\left|A_{11} \right|\left|A_{nn} \right|-\left|A_{1n} \right|\left|A_{n1} \right|.
\end{equation}
\end{Lemma}
We define the $k \times k$ matrix $X_k(n,t)$ as
\begin{equation}
X_k(n,t):=
\begin{pmatrix}
x_{n,t}&x_{n+2,t}&\cdots&x_{n+2k-2,t}\\
x_{n,t+1}&x_{n+2,t+1}&\cdots&x_{n+2k-2,t+1}\\
\vdots&\vdots&\ddots&\vdots\\
x_{n,t+k-1}&x_{n+2,t+k-1}&\cdots&x_{n+2k-2,t+k-1}
\end{pmatrix},
\end{equation} 
and its determinant $D_k(n,t)$ as $D_k(n,t):=| X_k(n,t) |$.
From here on let us use the notation that $P\equiv 0$ if $P$ is identically zero as a function of the initial variables of the given recurrence. 
\begin{Proposition}\label{prop_nonzero}
We have $x_{n,t} \not\equiv 0$, $D_2(n,t) \not\equiv 0$,
and 
$D_3(n,t) \not\equiv 0$.
\end{Proposition}
\Proof
If we assign positive values to all the initial data, then we have $x_{n,t}>0$.
Hence $x_{n,t} \not\equiv 0$. 
\[
D_2(n,t)=\begin{vmatrix}
x_{n,t}&x_{n+2,t}\\
x_{n,t+1}&x_{n+2,t+1}
\end{vmatrix}=x_{n+1,t}+x_{n+1,t+1} \not\equiv 0.
\]
By choosing the initial condition so that $x_{n,t}=x_{n+1,t-1}$,
the sequence $\{a_m\}$ with $a_m:=x_{n,t},\;m:=n+t$ satisfies
\[
a_{m+3}a_m=a_{m+2}a_{m+1}+a_{m+1}+a_{m+2},
\]
which is equivalent to the Heideman-Hogan recurrence \eqref{HH_eq} with $k=1$.
From Lemma \ref{Dodgson} we have
\[
a_{m+3} D_3(n,t)=(a_{m+4}+a_{m+5})(a_{m+1}+a_{m+2})-(a_{m+3}+a_{m+4})(a_{m+2}+a_{m+3}).
\]
Therefore
\[
D_3(n,t)=\frac{a_{m+2}+2a_{m+3}+a_{m+4}+a_{m+1}a_{m+5}-a_{m+3}^2}{a_{m+3}}.
\]
If we take all the initial values as positive, then $a_m>0$ for any $m$.
Since we have
\[
\frac{a_{m+3}}{a_{m+1}}=\frac{a_{m+2}}{a_m}+\frac{a_{m+1}+a_{m+2}}{a_ma_{m+1}}>\frac{a_{m+2}}{a_m},
\]
$a_{m+1}a_{m+5}-a_{m+3}^2>0$, which implies $D_3(n,t)>0$.
Therefore $D_3(n,t)\not\equiv 0$.
\qed

\begin{Proposition}\label{prop_3by3}
The determinant $D_3(n,t)$ does not depend on $n$:
\begin{equation}\label{linear_equality}
D_3(n,t)=D_3(n+1,t).
\end{equation}
\end{Proposition}
Proof of proposition \ref{prop_3by3} is in the appendix.
\begin{Corollary}\label{cor1}
\begin{equation}\label{D4eqzero}
D_4(n,t)\equiv 0
\end{equation}
\end{Corollary}
\Proof
From Lemma \ref{Dodgson}, we have
\[
D_2(n+2,t+1)D_4(n,t)=D_3(n,t)D_3(n+2,t+1)-D_3(n+2,t)D_3(n,t+1).
\]
Proposition \ref{prop_3by3} implies that $D_3(n,t)=D_3(n+2,t)$ and that $D_3(n,t+1)=D_3(n+2,t+1)$.
Since $D_2(n+2,t+1)\not\equiv 0$ from Proposition \ref{prop_nonzero}, we have $D_4(n,t)\equiv 0$.
\qed
\hbreak
\textbf{Proof of Theorem \ref{Theorem_main}}

Proof of \eqref{Lin1}:\ 
From $D_4(n,t)=0$, $X_4(n,t)$ has a null right eigenvector $\p(n,t)$:
\[
X_4(n,t)\p(n,t)=X_4(n,t+1)\p(n,t+1)=\b0.
\]
Since $X_4(n,t)$ and $X_4(n,t+1)$ have the same three rows in common, and both of their ranks are three from Proposition \ref{prop_nonzero}, two vectors $\p(n,t)$ and $\p(n,t+1)$ are linearly dependent.
Hence we can choose a null right eigenvector $\p(n)$ of $X_4(n,t)$
independent of $t$.
Furthermore, from the cofactor expansion,
\[
D_4(n,t)=D_3(n+2,t+1)x_{n,t}-D_3^{(12)}(n,t)x_{n+2,t}+D_3^{(13)}(n,t)x_{n+4,t}-D_3(n,t+1)x_{n+6,t}=0,
\]
where $D_3^{(ij)}(n,t)$ is the $(ij)$ minor of $X_4(n,t)$.
Therefore one of the null right eigenvectors is
\[
{}^{t} (D_3(n+2,t+1),-D_3^{(12)}(n,t),D_3^{(13)}(n,t),-D_3(n,t+1))
\]
Since $D_3(n,t+1)=D_3(n+2,t+1)$, we can choose
\[
\p(n)={}^t (-1,\beta(n), \alpha(n),1).
\] 
Similarly, by considering a null left eigenvector $\q(n,t)$ of $X_4(n,t)$, we have
\[
\q(n,t) X_4(n,t)=\q(n+2,t)X_4(n+2,t)=0.
\]
By the same argument as before, $\q(n,t)$ does not depend on $n$ except for its parity.
We obtain \eqref{Lin2}
and \eqref{Lin3} in the same manner as in \eqref{Lin1}.
\qed
\hbreak
Let $b_n:=x_{n,t=0}$, then the following proposition shows the explicit form of $\alpha(n)$, $\beta(n)$.
\begin{Proposition}\label{Coefficients}
The coefficients $\alpha(n)$ and $\beta(n)$ ($n=0,1,2,...$) of \eqref{Lin1} are given as
\begin{subequations}
\begin{align}
\alpha(n)&=-\frac{1+b_{n+1}b_{n+4}+b_{n+2}b_{n+5}+b_{n+3}b_{n+6}}{b_{n+3}b_{n+4}}\label{alpha_n}\\
\beta(n)&=\frac{1+b_{n}b_{n+3}+b_{n+1}b_{n+4}+b_{n+2}b_{n+5}}{b_{n+2}b_{n+3}}\label{beta_n}
\end{align}
\end{subequations}
In particular, we find $\alpha(n)=-\beta(n+1)$ for $n\ge 0$.
\end{Proposition}
Proof of proposition \ref{Coefficients} is straightforward and is found in the appendix.
\begin{Proposition} \label{propofpqrs}
Let $p_0:=x_{0,t}$, $p_1:=x_{0,t+1}$, $p_2:=x_{0,t+2}$, $p_3:=x_{0,t+4}$ and $q_0:=x_{1,t}$, $q_1:=x_{1,t+1}$, $q_2:=x_{1,t+2}$, $q_3:=x_{1,t+4}$.
The variables $\gamma(t)$, $\ldots$, $\epsilon'(t)$ in \eqref{Lin2} and \eqref{Lin3}  are given as
\begin{subequations}
\begin{align}
\gamma(t)&=-D\left(p_1p_2q_0q_1 + p_1p_3q_0q_1 + p_1p_2q_1^2 + p_1p_3q_1^2 + p_1^2q_0q_2 + p_1p_2q_0q_2+ p_1p_3q_0q_2\right.\nonumber \\
&\quad +p_2p_3q_0q_2+p_1^2q_1q_2 +p_1p_2q_1q_2+p_0p_3q_1q_2 +2 p_1p_3q_1q_2+ p_2p_3q_1q_2+p_0p_1q_2^2 + p_1^2q_2^2  \nonumber\\
&\quad \left. +  p_0p_3q_2^2 + p_1p_3q_2^2 + p_1^2q_0q_3+ p_1p_2q_0q_3 + p_1^2q_1q_3  + p_1p_2q_1q_3 + p_0p_1q_2q_3 + p_1^2q_2q_3 \right), \label{tlin1}\\
\delta(t)&=D\left(p_2^2q_0q_1 + p_2p_3q_0q_1 + p_0p_2q_1^2 + p_2^2q_1^2 + p_0p_3q_1^2 + 
p_2p_3q_1^2 + p_1p_2q_0q_2+ p_2^2q_0q_2 \right.\nonumber\\
&\quad + p_0p_1q_1q_2 + 2p_0p_2q_1q_2 +
p_1p_2q_1q_2+p_2^2q_1q_2+p_0p_3q_1q_2+p_0p_2q_2^2+p_1p_2q_2^2+p_1p_2q_0q_3\nonumber\\
&\quad + \left. p_2^2q_0q_3+p_0p_1q_1q_3+p_0p_2q_1q_3+p_1p_2q_1q_3+ 
p_2^2q_1q_3+p_0p_2q_2q_3+p_1p_2q_2q_3\right), \label{tlin2}\\
\epsilon(t)&=-p_1q_1D\left(p_2q_1+p_3q_1+p_1q_2+2p_2q_2+p_3q_2+p_1q_3+p_2q_3\right), \label{tlin3}\\
\gamma'(t)&=-D \left( p_2^2q_0q_1 +p_2p_3q_0q_1 +p_0p_2q_1^2 +p_1p_2q_1^2 +p_2^2q_1^2 +p_0p_3q_1^2 +p_1p_3q_1^2 +p_2p_3q_1^2 \right.\nonumber\\
&\quad +p_0p_1q_1q_2 +p_1^2q_1q_2 +p_0p_2q_1q_2 +p_1p_2q_1q_2 +p_0p_3q_1q_2 +p_1p_3q_1q_2 +p_1p_2q_0q_3 +p_2^2q_0q_3 \nonumber\\
&\quad +\left.p_0p_1q_1q_3 +p_1^2q_1q_3 +p_0p_2q_1q_3 +2p_1p_2q_1q_3 +p_2^2q_1q_3 +p_0p_2q_2q_3 +p_1p_2q_2q_3\right), \label{tlin4}\\
\delta'(t)&=D\left(p_1p_2q_0q_1 +p_1p_3q_0q_1 +p_1^2q_0q_2 +2p_1p_2q_0q_2+p_2^2q_0q_2 + 
p_1p_3q_0q_2 +p_2p_3q_0q_2\right.\nonumber\\
&\quad +p_0p_2q_1q_2+p_1p_2q_1q_2+p_2^2q_1q_2 + 
p_0p_3q_1q_2 +p_1p_3q_1q_2 +p_2p_3q_1q_2 +p_0p_1q_2^2 +p_1^2q_2^2 \nonumber \\
&\quad \left. +p_0p_2q_2^2+p_1p_2q_2^2+p_0p_3q_2^2 +p_1p_3q_2^2 +p_1^2q_0q_3 +
p_1p_2q_0q_3 +p_0p_1q_2q_3 +p_1^2q_2q_3 \right), \label{tlin5}\\
\epsilon'(t)&= \epsilon(t), \label{tlin6}
\end{align}
\end{subequations}
where 
$D^{-1}=p_2q_2(p_1q_0+p_2q_0+p_0q_1+2p_1q_1+p_2q_1+p_0q_2+p_1q_2)$.
\end{Proposition}
Proof of proposition \ref{propofpqrs} is in the appendix.
%
%
\section{Reduction, Laurent property and algebraic Entropy}
In this section, we consider a reduction of \eqref{2DHH}.
Let $K,\, M$ be mutually co-prime positive integers.
We impose the constraint:
\begin{equation}\label{reduction_MK}
x_{n,t}=x_{n+M,t-K}.
\end{equation}
For $j=nK+tM$, we put $a_j:=x_{n,t}$. 
Then $\{a_j\}$ satisfies the nonlinear recurrence:
\begin{equation}\label{KMreduction}
a_{j+2K+M}a_j=a_{j+M}a_{j+2K}+a_{j+M+K}+a_{j+K},
\end{equation}
which is a generalization of the Heideman-Hogan recurrence \eqref{HH_eq}.
\begin{Proposition}\label{prop_coeff}
Let us consider the linear equations \eqref{Lin1}, \eqref{Lin2}, \eqref{Lin3} corresponding to \eqref{2DHH}
with the reduction \eqref{reduction_MK}. We have 
\begin{equation}\label{n_difec_period}
\alpha(n+M)=\alpha(n), \quad \beta(n+M)=\beta(n).
\end{equation}
If $M$ is an odd integer,
\begin{equation}\label{t_direc_period1}
\left\{
\begin{array}{lll}
\gamma(t+K)=\gamma'(t),\quad &\delta(t+K)=\delta'(t),\quad \epsilon(t+K)=\epsilon'(t),\\
\gamma'(t+K)=\gamma(t),\quad &\delta'(t+K)=\delta(t),\quad \epsilon'(t+K)=\epsilon(t),
\end{array}
\right.
\end{equation}
while if $M$ is an even integer,
\begin{equation}
\left\{
\begin{array}{lll}
\gamma(t+K)=\gamma(t),\quad &\delta(t+K)=\delta(t),\quad &\epsilon(t+K)=\epsilon(t),\\
\gamma'(t+K)=\gamma'(t),\quad &\delta'(t+K)=\delta'(t),\quad &\epsilon'(t+K)=\epsilon'(t).
\end{array}
\right.
\label{t_direc_period2}
\end{equation}
\end{Proposition}
\Proof
In the proof of Theorem \ref{Theorem_main}, we notice
\[
\alpha(n)=-\frac{D_3^{(13)}(n,t)}{D_3(n,t+1)}.
\]
It is easy to see that $D_3^{(13)}(n,t)=D_3^{(13)}(n+M,t-K)$, $D_3(n,t+1)=D_3(n+M,t-K+1)$ and therefore $\alpha(n)=\alpha(n+M)$.
In a similar manner we obtain
\[
\gamma(n)=-\frac{D_3^{(31)}(n,t)}{D_3(n+2,t)},
\]
and from this relation we have
\[
\gamma(t+K)=
\left\{
\begin{array}{ll}
\gamma'(t) & (M\ \text{is odd})\\
\gamma(t) & (M\ \text{is even})
\end{array}
\right.
.
\]
Other periodicities can be obtained in similar manners.
\qed
\hbreak
From Theorem \ref{Theorem_main}, we have the following corollary:
\begin{Corollary}\label{cor_linear_redeq1}
Let $\tilde{\alpha}(j)=\alpha(n),\, \tilde{\beta}(j)=\beta(n)$, then 
$\tilde{\alpha}(j+M)=\tilde{\alpha}(j),\, \tilde{\beta}(j+M)=\tilde{\beta}(j)$ and we have a linear equation
\begin{equation}\label{linear_red_eq1}
a_{j+6K}+\tilde{\alpha}(j)a_{j+4K}+\tilde{\beta}(j)a_{j+2K}-a_j=0.
\end{equation}
We also define
\begin{equation}\label{gdef_M_even}
\tilde{\gamma}(j)=\left\{
\begin{array}{cl}
\gamma(t) &(j=2mK+tM)\\
\gamma'(t)&(j=(2m+1)K+tM)
\end{array}
\right.
\end{equation}
It holds that $\tilde{\gamma}(j+2K)=\tilde{\gamma}(j)$. The terms $\tilde{\delta}(j)$ and $\tilde{\epsilon}(j)$ are also defined in the same manner and have the same periodicity relations.
Then, we have a linear equation
\begin{equation}\label{linear_red_eq2}
a_{j+3M}+\tilde{\gamma}(j)a_{j+2M}+\tilde{\delta}(j)a_{j+M}+\tilde{\epsilon}(j)a_j=0.
\end{equation}
\end{Corollary}
To prove the corollary \ref{cor_linear_redeq1}, roughly speaing, we need to show that the coefficients of the reduced equations are well-defined from the periodicity. The details are found in the appendix.
\begin{Proposition}\label{prop_red_coefficients}
The functions $\tilde{\alpha}(j)$ and $\tilde{\beta}(j)$ are given by：
\begin{subequations}
\begin{align}
\tilde{\alpha}(j)&=-\frac{1+a_{(j+1)K}a_{(j+4)K}+a_{(j+2)K}a_{(j+5)}+a_{(j+3)K}a_{(j+6)K}}{a_{(j+3)K}a_{(j+4)K}}\label{talpha_j},\\
\tilde{\beta}(j)&=\frac{1+a_{jK}a_{(j+3)K}+a_{(j+1)K}a_{(j+4)K}+a_{(j+2)K}a_{(j+5)K}}{a_{(j+2)K}a_{(j+3)K}}\label{tbeta_j},
\end{align}
\end{subequations}
In particular，$\tilde{\alpha}(j)=-\tilde{\beta}(j+K)$.
\end{Proposition}
\Proof
This proposition is a direct consequence of Proposition \ref{Coefficients}. 
\qed
\hbreak
Similarly, we have explicit expressions for $\tilde{\gamma}(j)$ and so on, however we omit the details here.
The next proposition is due to \cite{Gnanadoss}, and is proved using the Cayley-Hamilton theorem in linear algebras.
\begin{Proposition}\label{linear_eq}
The sequence $(a_j)$ satisfies a linear recurrence of order $6KM$ with constant coefficients.
\end{Proposition}
At the latter half of this section, we shall prove that the iterate $x_{n,t}$ of \eqref{2DHH} is an irreducible Laurent polynomial of the initial variables, and that every pair of two iterates are coprime as Laurent polynomials.
For simplicity let us study the evolution of \eqref{2DHH} over the first quadrant
from the initial variables at the three half-lines $t=0, n=0, n=1$.
Let us note that a similar discussion holds for other initial configurations such as the staircase one.
\begin{Theorem}
We have the Laurent  property:
\[
x_{n,t}\in R:=\mathbb{Z}\left[ \{x_{n,0}^{\pm}, x_{0,t}^{\pm}, x_{1,t}^{\pm} \}_{n,t\in\mathbb{N}} \right].
\]
The iterate $x_{n,t}$ is irreducible in $R$, and two distinct iterates are coprime in $R$.
\end{Theorem}
\Proof
Let us define
\[
R_{n,t}:=\mathbb{Z}\left[ \{x_{m,0}^{\pm}, x_{0,s}^{\pm}, x_{1,s}^{\pm} \}_{0\le m\le n, 0\le s\le t} \right].
\]
We prove the following three facts by an induction with respect to $n$ and $t$.
\begin{enumerate}
\item $x_{n,t}\in R_{n,t}$.
\item $x_{n,t}$ is irreducible in $R_{n,t}$.
\item $x_{n,t}$ is coprime with all $x_{m,s}$ with $0\le m\le n, 0\le s\le t, (m,s)\neq (n,t)$.
\end{enumerate}
Let us assume that the properties 1, 2 and 3 are all satisfied for every $(m,s)$ with
$m\le n$, $s\le t$, $(m,s)\neq (n,t)$, and prove these properties for $(m,s)=(n,t)$.
First let us prove $x_{n,t}\in R_{n,t}$. We calculate the right hand side of
\[
x_{n,t}x_{n-2,t-1}=x_{n-2,t}x_{n,t-1}+x_{n-1,t}+x_{n-1,t-1}=:F.
\]
We have $F\in R_{n,t}$ from the induction hypothesis.
To ease notation, we reassign the subscripts of the variables as $n-4\to 0$, $t-2\to 0$: e.g., $x_{42}:=x_{n,t}$, $x_{21}:=x_{n-2,t-1}$, and
\[
x_{42}=\frac{x_{22}x_{41}+x_{32}+x_{31}}{x_{21}}.
\]
Let us prove that $F$ vanishes modulo $x_{21}$.
Since
\[
x_{41}\equiv \frac{x_{31}+x_{30}}{x_{20}},\ x_{32}\equiv \frac{x_{12}x_{31}+x_{22}}{x_{11}}\mod x_{21},
\]
we have
\[
x_{32}+x_{31}\equiv \frac{x_{22}}{x_{11}}+\frac{x_{31}}{x_{11}}(x_{12}+x_{11})\equiv
\frac{x_{22}}{x_{11}}+\frac{x_{31}}{x_{11}}(x_{22}x_{01})\equiv \frac{x_{22}}{x_{11}}(1+x_{31}x_{01}),
\]
and
\begin{align*}
F&\equiv \frac{x_{22}}{x_{20}x_{11}}(x_{31}x_{11}+x_{30}x_{11}+x_{20}+x_{31}x_{01}x_{20})=\frac{x_{22}x_{31}}{x_{20}x_{11}}(x_{11}+x_{10}+x_{01}x_{20})\\
& \equiv \frac{x_{22}x_{31}}{x_{20}x_{11}} x_{00}x_{21}\equiv 0 \mod x_{21}.
\end{align*}
Therefore we have
\[
x_{n,t}x_{n-2,t-1}=\tilde{F} x_{n-2,t-1},
\]
where $\tilde{F}\in R$. Thus $x_{n,t}\in R$.

Next we shall prove the irreducibility of $x_{n,t}$.
Let us take $X:=x_{n,0}$.
\[
x_{n,t}=\frac{1}{x_{n-2,t-1}}\left\{ x_{n-2,t}x_{n,t-1}+x_{n-1,t}+x_{n-1,t-1} \right\}
\]
is linear with respect to $x_{n,t-1}$.
By continuing the iterations towards $t=0$, we have
\[
x_{n,t}=A_{n,t} X+B_{n,t},
\]
where $A_{n,t}$ and $B_{n,t}$ do not depend on $X$.
Let us prepare a lemma on the first order polynomial in $R$:
\begin{Lemma} \label{lemma2}
Let $g_1,g_2$ be non-zero Laurent polynomials in $R$. Suppose that $g_1,g_2$ are coprime in $R$ and that $g_1,g_2$ do not depend on a variable $x$.
Then $f:=xg_1+g_2$ is irreducible. Moreover, $f$ is coprime with every $g\in R$ which does not depend on $x$.
\end{Lemma}
From
\begin{align}
A_{n,t}&= \frac{x_{n-2,t}}{x_{n-2,t-1}}A_{n,t-1},\\
B_{n,t}&= \frac{1}{x_{n-2,t-1}}\left( B_{n,t-1}x_{n-2,t}+x_{n-1,t}+x_{n-1,t-1} \right).
\end{align}
we have
\[
A_{n,t}=\frac{x_{n-2,t}}{x_{n-2,0}},
\]
for all non-negative $t$. If we substitute $1$ for all the initial variables, we have $x_{n,t}>0$ for all $t$ and $n$. Therefore $A_{n,t}$ is not identically equal to zero.
In a similar manner, we can conclude that $B_{n,t}$ is not identically zero.
Next we prove the coprimeness of $A_{n,t}$ and $B_{n,t}$.
Let us suppose otherwise, then $x_{n-2,t}$ and $(x_{n-1,t-1}+x_{n-1,t})$ must have a non-monomial common factor. Since
\[
x_{n-1,t-1}+x_{n-1,t}=x_{n-2,t-1}x_{n,t}-x_{n-2,t}x_{n,t-1},
\]
$x_{n-2,t}$ and $(x_{n-2,t-1}x_{n,t})$ must share a non-monomial common factor.
Since $x_{n-2,t}$ and $x_{n-2,t-1}$ are mutually coprime from the induction hypothesis,
the two iterates $x_{n-2,t}$ and $x_{n,t}$ are not coprime. However, again from the induction hypothesis that $x_{n-2,t}$ is irreducible, $x_{n,t}$ must be divisible by $x_{n-2,t}$ in $R$. This is in contradiction with the fact that $x_{n,t}=(x_{n-1,t}+x_{n-1,t-1})/x_{n-2,t-1}\neq 0$ when $x_{n-2,t}=0$.
Therefore using Lemma \ref{lemma2} we have the irreducibility of $x_{n,t}$.

Finally we prove the property 3: the coprimeness.
From lemma \ref{lemma2}, $x_{n,t}$ is coprime with $x_{m,s}$ if
$m\le n-1, s\le t$.
We need to prove the coprimeness of $x_{n,t}$ with $x_{n,s}$ $(s=0,1,2,\cdots,t-1)$.
Let us suppose otherwise.
From the induction hypothesis that $x_{n,s}$ $(s\le t-1)$ is irreducible,
we have $x_{n,t}=M x_{n,s}$, where $M$ is a monomial of $R$.
Then, by substituting $1$ for all the initial variables, we have $M=1$ and thus
$x_{n,t}=x_{n,s}$, which is a contradiction to the strictly increasing property of the values.

\qed

Finally, we show that the degree of $x_{n,t}$ grows as a polynomial of the first order.
Let us decompose $x_{n,t}$ as
\[
x_{n,t}=\frac{p_{n,t}}{q_{n,t}},
\]
where $q_{n,t}$ is a monomial in $R$, and $p_{n,t}$, $q_{n,t}$ are mutually coprime polynomials.
\begin{Proposition}
Let us define
\[
L_{n,t}:=\text{LCM}\, (q_{n-2,t}q_{n,t-1}, q_{n-1,t}, q_{n-1,t-1}).
\]
Then
\begin{align*}
p_{n,t}& =\left( \frac{p_{n-2,t}p_{n,t-1}}{q_{n-2,t}q_{n,t-1} } +\frac{p_{n-1,t}}{q_{n-1,t}} +\frac{p_{n-1,t-1}}{q_{n-1,t-1}}\right) \frac{L_{n,t}}{p_{n-2,t-1}}, \\
q_{n,t}& = \frac{L_{n,t}}{q_{n-2,t-1}}.
\end{align*}
Moreover, the explicit form of $q_{n,t}$ expressed in terms of the initial variables is
\[
q_{n,t}=\left(\prod_{0\le m\le n-2} x_{m,0}\right) \left(\prod_{1\le s\le t-1}x_{0,s}x_{1,s}\right).
\]
\end{Proposition}
The proof is done using a discussion in \cite{Mase} and is omitted here.
From the expression of $q_{n,t}$, the degree of $q_{n,t}$ is $(2t+n-3)$, which exhibits the first order growth in both of the directions $n$ and $t$.
Note that $\deg p_{n,t}=\deg q_{n,t}+1$.
Therefore the algebraic entropy of the system \eqref{2DHH} is zero, and thus the system is integrable in its sense.

%
%
%
%
%
%
\section{Lattice equations related to Dana-Scott sequence}
Let us review linearizability of the equation for the two-frieze pattern\cite{M-GOT}
\begin{equation}\label{Toda_like}
x_{n,t}x_{n+2,t+2}=x_{n+2,t}x_{n,t+2}+x_{n+1,t+1}\quad ((n,t) \in (2\Z)^2 \mbox{\,or\,}
(2\Z+1)^2),
\end{equation}
which is equivalent to \eqref{two_frieze}.
We denote by $F_k(n,t)$ the determinant of the $k \times k$ matrix the $(ij)$ entry of which is $x_{n+2(i-1), t+2(j-1)}$: i.e.,
\begin{equation}\label{F_k-form}
F_k(n,t):=
\begin{vmatrix}
x_{n,t}&x_{n,t+2}&\ldots&x_{n,t+2(k-1)}\\
x_{n+2,t}&x_{n+2,t+2}&\ldots&x_{n+2,t+2(k-1)}\\ 
\vdots &\vdots &\ddots&\vdots \\
x_{n+2(k-1),t}&x_{n+2(k-1),t+2}&\ldots&x_{n+2(k-1),t+2(k-1)}
\end{vmatrix}.
\end{equation}
We also define $F_0(n,t):=1$.
From the Dodgson identity of Lemma \ref{Dodgson} and \eqref{Toda_like} we have
\begin{align*}
x_{n+2,t+2}F_3(n,t)
&=F_2(n,t)F_2(n+2,t+2)-F_2(n+2,t)F_2(n,t+2)\\
&=x_{n+1,t+1}x_{n+3,t+3}-x_{n+3,t+1}x_{n+1,t+3} \\
&=x_{n+2,t+2}.
\end{align*}
Hence we find
\[
F_3(n,t)=1.
\]
Using Lemma \ref{Dodgson} again,
\begin{align*}
F_2(n+2,t+2)F_4(n,t)&=F_3(n,t)F_3(n+2,t+2)-F_3(n+2,t)F_3(n,t+2)\\
&=1\cdot 1-1 \cdot 1 =0.
\end{align*}
Therefore we have
\[
F_4(n,t) \equiv \begin{vmatrix}
x_{n,t}&x_{n+2,t}&x_{n+4,t}&x_{n+6,t}\\
x_{n,t+2}&x_{n+2,t+2}&x_{n+4,t+2}&x_{n+6,t+2}\\
x_{n,t+4}&x_{n+2,t+4}&x_{n+4,t+4}&x_{n+6,t+4}\\
x_{n,t+6}&x_{n+2,t+6}&x_{n+4,t+6}&x_{n+6,t+6}
\end{vmatrix}
=0.
\]
Hence, from arguments similar to those in the Proof of Theorem \ref{Theorem_main}, we have linear equations
\begin{align*}
&x_{n+6,t}+K_1^{(+)}(n)x_{n+4,t}+K_2^{(+)}(n)x_{n+2,t}-x_{n,t}=0\quad \mbox{for $t \in 2\Z$},\\
&x_{n+6,t}+K_1^{(-)}(n)x_{n+4,t}+K_2^{(-)}(n)x_{n+2,t}-x_{n,t}=0\quad \mbox{for $t \in 2\Z+1$},\\
&x_{n,t+6}+L_1^{(+)}(t)x_{n,t+4}+L_2^{(+)}(t)x_{n,t+2}-x_{n,t}=0\quad \mbox{for $t \in 2\Z$},\\
&x_{n,t+6}+L_1^{(-)}(t)x_{n,t+4}+L_2^{(-)}(t)x_{n,t+2}-x_{n,t}=0\quad \mbox{for $t \in 2\Z+1$},
\end{align*}
where $K_1^{(\pm)},\,K_2^{(\pm)}$ (resp. $L_1^{(\pm)},\,L_2^{(\pm)})$ are functions of $n$ (resp. $t$) determined from the boundary and/or the initial conditions.  

Now we consider generalization of equation~\eqref{Toda_like}.
Natural extension of equation~\eqref{Toda_like} would be the followings  ($(n,t) \in (2\Z)^2, (2\Z+1)^2$) :
\begin{equation}\label{Det_like1}
F_3(n,t)=x_{n+2,t+2}
\end{equation}
or 
\begin{equation}\label{Det_like2}
F_3(n,t)=F_2(n+1,t+1)
\end{equation}
Both \eqref{Det_like1} and \eqref{Det_like2} are linearizable. 
In fact, using the same arguments, we have $F_5(n,t)=0$ for \eqref{Det_like1}, and
$F_6(n,t)=0$ for \eqref{Det_like2}.
Hence we find, for example,  
\[
x_{n,t}+\alpha(n)x_{n+2,t}+\beta(n)x_{n+4,t}+\gamma(n)x_{n+6,t}+\delta(n)x_{n+8,t}-x_{n+10,t}=0
\]
for \eqref{Det_like2}.
In general, we have the following proposition:
\begin{Proposition}\label{Det_general}
Let $k$ be a non-negative integer, and $F_k(n,t)$ be the determinant
defined by \eqref{F_k-form} $(F_0(n,t)=1)$.
If it holds that
\begin{equation}\label{Det_k2}
F_{k+1}(n,t)=F_k(n+1,t+1),
\end{equation}
then $F_{2k+2}=0$, and $x_{n,t}$ satisfy the linear equations
\begin{subequations}
\begin{align}
&x_{n,t}+C_1^{(\pm)}(n)x_{n+2,t}+C_2^{(\pm)}(n)x_{n+4,t}+\cdots+C_{2k}^{(\pm)}(n)x_{n+4k,t}-x_{n+4k+2,t}=0,\label{gcoef1}\\
&x_{n,t}+{C_1^{(\pm)}}'(t)x_{n,t+2}+{C_2^{(\pm)}}'(t)x_{n,t+4}+\cdots+{C_{2k}^{(\pm)}}'(t)x_{n,t+4k}-x_{n,t+4k+2}=0,\label{gcoef2}
\end{align}
\end{subequations}
where $C_i^{(+)}(n)$ and $C_i^{(-)}(n)$ $(i=1,2,...,2k)$ are functions of $n$ for even $t$ and odd $t$ respectively, and ${C_i^{(+)}}'(t)$ and ${C_i^{(-)}}'(t)$ $(i=1,2,...,2k)$ are functions of $t$ for even $n$ and odd $n$ respectively.

Similarly, if it holds that
\begin{equation}\label{Det_k1}
F_{k+2}(n,t)=F_k(n+2,t+2),
\end{equation}
then $F_{2k+3}=0$, and $x_{n,t}$ satisfy the linear equations
\begin{subequations}
\begin{align}
&x_{n,t}+S_1^{(\pm)}(n)x_{n+2,t}+S_2^{(\pm)}(n)x_{n+4,t}+\cdots+S_{2k+1}^{(\pm)}(n)x_{n+4k+2,t}+x_{n+4k+4,t}=0,\\
&x_{n,t}+{S_1^{(\pm)}}'(t)x_{n,t+2}+{S_2^{(\pm)}}'(t)x_{n,t+4}+\cdots+{S_{2k+1}^{(\pm)}}'(t)x_{n,t+4k+2}+x_{n,t+4k+4}=0,
\end{align}
\end{subequations}
where the notation $S_i^{(\pm)}(n)$ and ${S_i^{(\pm)}}'(t)$ $(i=1,2,...,2k+1)$ are interpreted in the same way as in \eqref{gcoef1} and \eqref{gcoef2}.
\end{Proposition}
Proof is straightforward and is found in the appendix.
%

By the reduction of the equations \eqref{Det_k2} and \eqref{Det_k1}, we have various linearizable recurrences of one variable.
For example, imposing $x_{n,t}=x_{n+1,t-1}$ on \eqref{Det_k2} for $k=2$ and putting $a_j=x_{j,j}$, we have
\begin{equation}\label{reduction2DHH}
a_{j+4}=\frac{a_ja_{j+3}^2+a_{j+2}^3+a_{j+1}a_{j+3}-2a_{j+1}a_{j+2}a_{j+3}-a_{j+2}^2}{a_{j+2}a_j-a_{j+1}^2}.
\end{equation}
The mapping \eqref{reduction2DHH} satisfies a 4th order linear equation,
however it does not have the Laurent property.
\section{Concluding remarks}
In this article, we have presented the linearizable equation \eqref{2DHH} defined over the two dimensional lattice.  
By imposing a constraint to the variable $x_{n,t}$, so that the equation is reduced on a line, it gives the Heideman-Hogan recurrence and its generalization.
We proved the Laurent property, the irreducibility and the coprimeness of \eqref{2DHH}. One comment is that the equation \eqref{2DHH} does not pass the singularity confinement test in the sense that its iterates always have monomial denominators, however, it passes the coprimeness criterion, whose concept is quite close to the singularity confinement but not exactly equivalent due to the different treatment of the monomial factors.
We have also studied the degree growth of \eqref{2DHH} and have proved that the degree grows as a first order polynomial with respect to $n$ and $t$. 
We then discussed a class of generalized lattice equations (generalization of \eqref{two_frieze}) corresponding to the two-frieze equation and the Dana-Scott recurrence.
The key lemma to construct these linearizable nonlinear equations is the Dodgson identity for matrix determinants.
Contrary to \eqref{two_frieze} and \eqref{2DHH}, the generalization \eqref{Det_k2} and \eqref{Det_k1} do not have the Laurent property.
However, since these equations are linearizable, they satisfy  \textit{extended} Laurent property.
For example, solutions $\{a_j\}$ to the recurrence \eqref{reduction2DHH} are polynomials in $\Z[a_1,a_2,a_3,a_4,(a_2^2-a_1a_3)^{-1},(a_3^2-a_2a_4)^{-1}]$. 
One of the future problems is to further investigate various classes of linearizable lattice equations and their properties including the extended Laurent property.
It is interesting to search for a suitable classification of the linearizable mappings defined on the two-dimensional lattice, for example, similarly to the classification of the one-dimensional linearizable equations into projective/Gambier/third-kind types.
\section*{Acknowledgement}
The authors are grateful to Prof. R. Willox for useful comments.
The present work is partially supported by KAKENHI Grant Numbers 16H06711 and 17K14211. 

\appendix
\section{Appendix: proof of propositions and corollaries}

\paragraph{Proof of proposition \ref{prop_3by3}}

From \eqref{2DHH} and Lemma \ref{Dodgson}, we have
\begin{align*}
x_{n+2,t+1}D_3(n,t)&=
\begin{vmatrix}
x_{n,t}&x_{n+2,t}\\
x_{n,t+1}&x_{n+2,t+1}
\end{vmatrix}     
\begin{vmatrix}
x_{n+2,t+1}&x_{n+4,t+1}\\
x_{n+2,t+2}&x_{n+4,t+2}
\end{vmatrix}
 -\begin{vmatrix}
x_{n+2,t}&x_{n+4,t}\\
x_{n+2,t+1}&x_{n+4,t+1}
\end{vmatrix}     
\begin{vmatrix}
x_{n,t+1}&x_{n+2,t+1}\\
x_{n,t+2}&x_{n+2,t+2}
\end{vmatrix}\\
&=\begin{vmatrix}
x_{n+1,t}+x_{n+1,t+1}&x_{n+3,t}+x_{n+3,t+1}\\
x_{n+1,t+1}+x_{n+1,t+2}&x_{n+3,t+1}+x_{n+3,t+2}
\end{vmatrix}.
\end{align*}
From the bilinearity of determinants, we have
\[
x_{n+2,t+1}D_3(n,t)=x_{n+2,t}+2x_{n+2,t+1}+x_{n+2,t+2}+x_{n+1,t}x_{n+3,t+2}-x_{n+1,t+2}x_{n+3,t}.
\]
By an up-shift of $n$, we have
\[
x_{n+3,t+1}D_3(n+1,t)=x_{n+3,t}+2x_{n+3,t+1}+x_{n+3,t+2}+x_{n+2,t}x_{n+4,t+2}-x_{n+2,t+2}x_{n+4,t}.
\]
Therefore,
\begin{align}
&x_{n+2,t+1}x_{n+3,t+1}\left(D_3(n,t)-D_3(n+1,t)\right)\nonumber\\
&=x_{n+3,t+1}(x_{n+2,t}+x_{n+2,t+2})-x_{n+2,t+1}(x_{n+3,t}+x_{n+3,t+2})\nonumber\\
&\qquad +
x_{n+3,t+1}(x_{n+1,t}x_{n+3,t+2}-x_{n+1,t+2}x_{n+3,t})\nonumber\\
&\qquad\qquad 
-x_{n+2,t+1}(x_{n+2,t}x_{n+4,t+2}-x_{n+2,t+2}x_{n+4,t}).
\label{mid_eq1}
\end{align}
By using Equation \eqref{Dodgson_id} to $x_{n+3,t+1}x_{n+1,t}$ and $x_{n+3,t+2}x_{n+1,t+1}$, we have two expressions for $x_{n+1,t+1}$ which read
\[
x_{n+1,t+1}=\frac{x_{n+1,t}x_{n+3,t+1}-x_{n+2,t}-x_{n+2,t+1} }{x_{n+3,t}}
=\frac{x_{n+3,t+1}x_{n+1,t+2}+x_{n+2,t+1}+x_{n+2,t+2} }{x_{n+3,t+2}}.
\]
From the second equality above we have
\begin{align}
&x_{n+3,t+1}(x_{n+1,t}x_{n+3,t+2}-x_{n+1,t+2}x_{n+3,t})\nonumber\\
&=
(x_{n+2,t}+x_{n+2,t+1})x_{n+3,t+2}+(x_{n+2,t+1}+x_{n+2,t+2})x_{n+3,t}.
\label{ideq1}
\end{align}
Similarly, by considering two expressions for $x_{n+4,t+1}$, we have
\begin{align}
&x_{n+2,t+1}(x_{n+2,t}x_{n+4,t+2}-x_{n+2,t+2}x_{n+4,t})\nonumber\\
&=
(x_{n+3,t}+x_{n+3,t+1})x_{n+2,t+2}+(x_{n+3,t+1}+x_{n+3,t+2})x_{n+2,t}
.\label{ideq2}
\end{align}
From \eqref{mid_eq1} and the identities \eqref{ideq1}, \eqref{ideq2}, 
\[
x_{n+2,t+1}x_{n+3,t+1}\left(D_3(n,t)-D_3(n+1,t)\right)=0.
\]
Since $x_{n+2,t+1}x_{n+3,t+1} \not\equiv 0$ by Proposition \ref{prop_nonzero},  we obtain the formula \eqref{linear_equality}.
\qed

\paragraph{Proof of proposition \ref{Coefficients}}

Let us denote $c_n:=x_{n,1}$.
From \eqref{2DHH}, we have
\begin{align*}
\frac{c_{n+6}}{b_{n+6}}-\frac{c_{n+4}}{b_{n+4}}=\frac{b_{n+5}+c_{n+5}}{b_{n+4}b_{n+6}},
\end{align*}
and thus
\begin{align*}
\frac{c_{n+6}}{b_{n+6}}-\frac{c_{n+2}}{b_{n+2}}
&=\frac{c_{n+6}}{b_{n+6}}-\frac{c_{n+4}}{b_{n+4}}+\frac{c_{n+4}}{b_{n+4}}-\frac{c_{n+2}}{b_{n+2}}\\
&=\frac{b_{n+5}+c_{n+5}}{b_{n+4}b_{n+6}}+\frac{b_{n+3}+c_{n+3}}{b_{n+2}b_{n+4}}.
\end{align*}
Furthermore we have
\begin{align*}
(b_{n+5}+c_{n+5})b_{n+3}&=(b_{n+3}+c_{n+3})b_{n+5}+(b_{n+4}+c_{n+4})\\
(b_{n+4}+c_{n+4})b_{n+2}&=(b_{n+2}+c_{n+2})b_{n+4}+(b_{n+3}+c_{n+3})\\
(b_{n+3}+c_{n+3})b_{n+1}&=(b_{n+1}+c_{n+1})b_{n+3}+(b_{n+2}+c_{n+2}),
\end{align*}
which yields
\[
b_{n+3}\left\{b_{n+2}(b_{n+5}+c_{n+5})+b_{n+4}(b_{n+1}+c_{n+1})   \right\}
=\left(1+b_{n+1}b_{n+4}+b_{n+2}b_{n+5} \right)(b_{n+3}+c_{n+3}).
\]
Equation \eqref{Lin1} for $t=0$ and $t=1$ yield
\begin{align*}
\begin{pmatrix}
\alpha(n)\\
\beta(n)
\end{pmatrix} 
&=
\begin{pmatrix}
c_{n+4} & c_{n+2}\\
b_{n+4} & b_{n+2}
\end{pmatrix}^{-1}
\begin{pmatrix}
c_n-c_{n+6}\\
b_n-b_{n+6}
\end{pmatrix}\\
&=\frac{1}{b_{n+3}+c_{n+3}}
\begin{pmatrix}
-(b_{n+1}+c_{n+1})-b_{n+2}c_{n+6}+b_{n+6}c_{n+2}\\
(b_{n+5}+c_{n+5})-b_{n+4}c_n+b_nc_{n+4}
\end{pmatrix}.
\end{align*}
Hence
\begin{align*}
\alpha(n)&=\frac{1}{b_{n+3}+c_{n+3}}\left\{ -(b_{n+1}+c_{n+1})-b_{n+2}b_{n+6}\left(\frac{c_{n+6}}{b_{n+6}}-\frac{c_{n+2}}{b_{n+2}} \right)\right\}\\
&=\frac{1}{b_{n+3}+c_{n+3}}\left\{ -(b_{n+1}+c_{n+1})-b_{n+2}b_{n+6}\left(\frac{b_{n+5}+c_{n+5}}{b_{n+4}b_{n+6}}+\frac{b_{n+3}+c_{n+3}}{b_{n+2}b_{n+4}} \right)\right\}\\
&=-\frac{ b_{n+2}(b_{n+5}+c_{n+5})+b_{n+4}(b_{n+1}+c_{n+1})  +b_{n+6}(b_{n+3}+c_{n+3}) }{(b_{n+3}+c_{n+3})b_{n+4}}\\
&=-\frac{ \left(1+b_{n+1}b_{n+4}+b_{n+2}b_{n+5} \right)(b_{n+3}+c_{n+3}) +b_{n+3}b_{n+6}(b_{n+3}+c_{n+3}) }{(b_{n+3}+c_{n+3})b_{n+3}b_{n+4}}\\
&=-\frac{1+b_{n+1}b_{n+4}+b_{n+2}b_{n+5}+b_{n+3}b_{n+6}}{b_{n+3}b_{n+4}}.
\end{align*}
Thus we obtain \eqref{alpha_n}.
Equation \eqref{beta_n} is obtained in a similar manner.
\qed

\paragraph{Proof of proposition \ref{propofpqrs}}
Let $r_i:=x_{2,t+i},\,y_i:=x_{4,t+i}$ ($i=0,1,2,...$).
We have simultaneous equations
\[
\begin{pmatrix}
p_0&p_{1}&p_{2}\\
r_0&r_{1}&r_{2}\\
y_0&y_{1}&y_{2}
\end{pmatrix}
\begin{pmatrix}
\epsilon(t)\\
\delta(t)\\
\gamma(t)
\end{pmatrix}
=-
\begin{pmatrix}
p_{3}\\
r_{3}\\
y_{3}
\end{pmatrix}.
\]
Hence we find
\[
\epsilon(t)=-\frac{1}{\det(t)}\begin{vmatrix}
p_3&p_{1}&p_{2}\\
r_3&r_{1}&r_{2}\\
y_3&y_{1}&y_{2}
\end{vmatrix},\;\;
\delta(t)=-\frac{1}{\det(t)}\begin{vmatrix}
p_0&p_{3}&p_{2}\\
r_0&r_{3}&r_{2}\\
y_0&y_{3}&y_{2}
\end{vmatrix},\;\;
\gamma(t)=-\frac{1}{\det(t)}\begin{vmatrix}
p_0&p_{1}&p_{3}\\
r_0&r_{1}&r_{3}\\
y_0&y_{1}&y_{3}
\end{vmatrix},
\]
where 
\[
\det(t)=\begin{vmatrix}
p_0&p_{1}&p_{2}\\
r_0&r_{1}&r_{2}\\
y_0&y_{1}&y_{2}
\end{vmatrix}.
\]
By using \eqref{2DHH} repeatedly, after a little tedious calculations, we prove the equations \eqref{tlin1} -- \eqref{tlin3}, and
\[
\det(t)=\frac{1}{p_1q_1}(p_1q_0+p_2q_0+p_0q_1+2p_1q_1+p_2q_1+p_0q_2+p_1q_2).
\]
Equations \eqref{tlin4} -- \eqref{tlin6} are calculated similarly from
\[
\begin{pmatrix}
q_0&q_{1}&q_{2}\\
s_0&s_{1}&s_{2}\\
z_0&z_{1}&z_{2}
\end{pmatrix}
\begin{pmatrix}
\epsilon'(t)\\
\delta'(t)\\
\gamma'(t)
\end{pmatrix}
=-
\begin{pmatrix}
q_{3}\\
s_{3}\\
z_{3}
\end{pmatrix},
\]
where we have used the notations $s_i:=x_{3,t+i}, z_i:=x_{5,t+i}$ $(i=0,1,2,\cdots)$.
\qed

\paragraph{Proof of corollary \ref{cor_linear_redeq1}}

Since $K$ and $M$ are mutually co-prime,
any integer $j$ can be expressed as
$j=nK+tM$ with an appropriate pair of integers $(n,t)$.
Let us take two expressions for $j$: $j=nK+tM=n'K+t'M$, then, 
$n \equiv n'$ (mod $M$) and $t \equiv t'$ (mod $K$).
Therefore $\alpha(n)=\alpha(n')$ by \eqref{n_difec_period}.
Thus $\tilde{\alpha}(j):=\alpha(n)$ is well-defined for every $j\in\mathbb{Z}$.
Since $j+M=nK+(t+1)M$, we have $\tilde{\alpha}(j+M)=\alpha(n)=\tilde{\alpha}(j)$.
The discussion for $\tilde{\beta}$ is the same.
Equation \eqref{linear_red_eq1} follows from \eqref{Lin1}.

\noindent
(i) In the case of even $M$:
Let us assume that $j$ has two decompositions $j=nK+tM=n'K+t'M$.
Since $M$ is even, $n\equiv n'\mod 2M$, and thus the parities of $n$ and $n'$ must coincide with each other. From \eqref{t_direc_period2} both $\gamma$ and $\gamma'$ have periods $K$. Therefore $\tilde{\gamma}(j)$ is well-defined.

\noindent
(ii) If $M$ is an odd integer, there exists an integer $i$ such that $n'=n+iM$ and $t'=t-iK$.
Since $\gamma(t-iK)=\gamma'(t)$ ($i$: odd) and $\gamma(t-iK)=\gamma(t)$ ($i$: even), we find that $\tilde{\gamma}(nK+tM)=\tilde{\gamma}(n'K+t'M)$, and thus, $\tilde{\gamma}(j)$ is well defined.

For $j=nK+tM$, $\tilde{\gamma}(j+2K)=\tilde{\gamma}((n+2)K+tM)=\tilde{\gamma}(nK+tM)$ and therefore $\tilde{\gamma}(j+2K)=\tilde{\gamma}(j)$.
The properties of $\tilde{\delta}(j)$ and $\tilde{\epsilon}(j)$ are proved similarly.
Finally, \eqref{Lin2} and \eqref{Lin3} give \eqref{linear_red_eq2} .
\qed

\paragraph{Proof of proposition \ref{Det_general}}
Let us prove
\begin{equation}
F_{k+i}(n,t)=F_{k-i+1}(n+2i-1,t+2i-1) \quad (i=1,2,...,k+1), \label{fkieq}
\end{equation}
by induction.
Suppose that \eqref{fkieq} holds for every positive integer $i=m$ ($1\le m \le k$).
Then, from Lemma \ref{Dodgson}, 
\begin{align*}
&F_{k+m-1}(n+2,t+2)F_{k+m+1}(n,t)\\
&=F_{k+m}(n,t)F_{k+m}(n+2,t+2)-F_{k+m}(n+2,t)F_{k+m}(n,t+2)\\
&=F_{k-m+1}(n+2m-1,t+2m-1)F_{k-m+1}(n+2m+1,t+2m+1)\\
&\quad -F_{k-m+1}(n+2m+1,t+2m-1)F_{k-m+1}(n+2m-1,t+2m+1)\\
&=F_{k-m}(n+2m+1,t+2m+1)F_{k-m+2}(n+2m-1,t+2m-1)\\
&=F_{k-m}(n+2m+1,t+2m+1)F_{k+m-1}(n+2,t+2).
\end{align*}
Hence we have $F_{k+m+1}(n,t)=F_{k-m}(n+2m+1,t+2m+1)$: i.e.,  \eqref{fkieq} is true for $i=m+1$.
For $i=1$, \eqref{fkieq} is nothing but \eqref{Det_k2}.
Thus \eqref{fkieq} is proved.
In particular, for $i=k+1$ we have
\[
F_{2k+1}(n,t)=F_0(n+2k+1,t+2k+1)=1,
\]
and, using Lemma \ref{Dodgson} again, we have $F_{2k+2}(n,t)=0$. 
Arguments similar to those in the proof of Theorem \ref{Theorem_main} lead \eqref{gcoef1} and \eqref{gcoef2}.
In the case of \eqref{Det_k1} can be done in the same manner and  is omitted. \qed


\begin{thebibliography}{99}

\bibitem{SC}
B. Grammaticos and A. Ramani and V. Papageorgiou,
Do integrable mappings have the
Painlev\'e property?,
Phys. Rev. Lett.,
\textbf{67},
1825--1828 (1991).

\bibitem{HV}
  J. Hietarinta and C. Viallet,
Singularity confinement and chaos in discrete systems,
  Phys. Rev. Lett.,  \textbf{81},  325--328 (1998).

\bibitem{RGH}
 A. Ramani and B. Grammaticos and J. Hietarinta,
Discrete versions of the Painlev\'e equations,
    Phys. Rev. Lett.,
   \textbf{67},
    1829--1832 (1991).
    
\bibitem{BV}
M. P. Bellon and C. M. Viallet,
Algebraic entropy,
Comm. Math. Phys. , \textbf{204},
425--437 (1999).

\bibitem{KMMT2}
M. Kanki, J. Mada, T. Mase and T. Tokihiro, 
Irreducibility and co-primeness as an integrability criterion for discrete equations
J. Phys. A, \textbf{47}, 465204 (2014).

\bibitem{KMT}
M. Kanki, T. Mase and T. Tokihiro,
Singularity confinement and chaos in two-dimensional discrete systems
J. Phys. A, \textbf{49}, 23LT01 (2015).

\bibitem{KKMT}
R. Kamiya, M. Kanki, T. Mase and T. Tokihiro,
Coprimeness-preserving non-integrable extension to the two-dimensional discrete Toda lattice equation,
\textit{J. Math. Phys.}, \textbf{58}, 012702 (2017).

\bibitem{RGSM}
A. Ramani, B. Grammaticos, J. Satsuma and N. Mimura,
Linearizable QRT mappings,
J. Phys. A: Math. Theor. \textbf{44}, 425201 (18pp)  (2011).

\bibitem{DS}
D. Gale,
The Strange and Surprising Saga of the Somos Sequences,
Mathematical Intelligencer,
\textbf{13},
40-42  (1991).

\bibitem{FZ}
S. Fomin and A. Zelevinsky,
Cluster algebras I: Foundations,
J. Amer. Math. Soc.,
\textbf{15}
497--529 (2002).

\bibitem{Fordy-Hone}
A. P. Fordy and A. Hone,
Discrete Integrable Systems and Poisson Algebras From Cluster Maps,
Comm. Math. Phys., 325, (2014), 527--584.

\bibitem{Propp}
J. Propp,
The combinatorics of frieze patterns and Markoff numbers,
arXiv:math/0511633.

\bibitem{M-GOT}
S. Morier-Genoud, V. Ovsienko and S. Tabachnikov,
2-Frieze patterns and the cluster structure of the space of
polygons, Ann. Inst. Fourier 62 (2012), 937-987.


\bibitem{Heideman-Hogan}
P. Heideman and E. Hogan, A New Family of Somos-like Recurrences, Electron.
J. Combin. 15 (2008) \#\!\! R54, 8pp.


\bibitem{FZ2}
S. Fomin and A. Zelevinsky,
The Laurent Phenomenon,
Adv. Applied Math., \textbf{28}, 119--144 (2002).


\bibitem{Hone-Ward}
Andrew N.W. Hone and Chloe Ward,
On the general solution of the Heideman-Hogan family of recurrences,
arXiv:1610.07199.


\bibitem{Dodgson}
C. L. Dodgson, 
Condensation of Determinants, Being a New and Brief Method for Computing their Arithmetical Values, Proc. R. Soc. Lond., 15 (1866), 150--155.


\bibitem{Gnanadoss}
A. A. Gnanadoss, Linear difference equations with periodic
coefficients, Proc. Amer. Math. Soc. 2 (1951), 699--703.

\bibitem{Mase}
T. Mase,
Investigation into the role of the Laurent property in integrability,
J. Math. Phys. 57 (2016), 022703.


\end{thebibliography}
\end{document}